# An Easy Cellular Gateway for Providing Shared Services and Data

Tayeb Lemlouma[1], Julien Luciana[2]
[1]IRISA Lab / University of Rennes I, [2]Centile
[1]Lannion, [2]Sophia Antipolis, France
Tayeb.Lemlouma@irisa.fr, jluciana@centile.com

Bastien Oza[3], Leandro Sierra[4], Mikaël Salun[5]
[3]Logicielnet, [4]University of Avignon, [5]University of Rennes I
bastien.oza@logicielnet.com, leandrosierra1@gmail.com, mikael.salaun@univ-rennes1.fr

*Abstract*— **In this paper, we present a new framework that links the two worlds of wired and cellular users sharing systems. The approach is to propose an easy gateway that enables the use of cellular networks based services by wireline users and applications. The idea is to use a mobile terminal or wireless equipment for sharing cellular services, available thanks to its cellular network, to other users that use the wireline Internet. The software application acts as a gateway between the cellular and the wired network; it is responsible for supporting the services provided by the wireless network and make them accessible and usable, in a standard and easy way, by anyone on the wireline network. The gateway software can be integrated easily on any complex architecture since it can interact with any cellular modem. The paper describes an implementation prototype where some examples of services, such as the ability of using messaging services and calls streaming, are experimented. The proposed platform combines different standards to guarantee the use of our gateway in heterogeneous environments.**

## I. INTRODUCTION

During last years, peer-to-peer (P2P) networks and sharing systems gained tremendous attention and popularity within both industry and academia. In wireline networks, content sharing environments enable Internet users to share any kind of content in a simple and free way. On the other hand, the use of small and mobile communications devices such as mobiles phones, smart phones and PDA's has gained momentum in our life. P2P computing and networking has been thoroughly studied in the literature either among classical users or mobile users (mobile P2P) [1][2][3][4][5]. If sharing data has made successes in wireline networks and generates a traffic that has been dominating the Internet, implementing P2P within wireless environments still faced to many challenges: starting from the question about the real need of sharing content that is often self created and privately consumed [6] and ending with the difficulties of implementing mobile P2P services. Such limitations are the complexity of cellular network architectures and the limitations of the bandwidth especially for uplink even within 3G networks. Moreover, even for wired or cellular networks, most of P2P applications are only single-service based where generally the only capability is the popular file sharing.

This paper describes a gateway framework for building mobile information sharing services. We propose an easy gateway based on cellular network services. The basic elements of the proposed architecture is composed of: 1) a wireless modem supporting 2G to 3G networks, 2) a set of exploitable services and 3) a software application that makes these services shared between the mobile terminal and users connected through the wireline network. The wireless modem can be used, e.g., by the mean of a simple mobile phone or a personal assistant device but also by the mean of professional equipment such as a cellular modem. One of the fixed requirements was that such gateway must be easy to integrate, do not need a high significant investment for the end user or application and could be used in heterogeneous environments. The objective of the implemented services, offered by the gateway, is to make them controlled and accessible for users whatever their locations. This enables users of the wired network to gain access to typically cellular services such as joining other mobile users using SMS/MMS messaging. Another use case of our gateway is to make a user's own data accessible from everywhere and anyhow even using a simple Internet connection. Data could be also shared between users exactly like in a classical P2P networks. Here, the particularity is that the data come from the mobile terminal and the wired (not the cellular) network is used for data sharing unlike mobile peer to peer approaches. Web services are used to make the gateway exploitable by any application in a standard way. Such applications could, for instance, offer online messaging services and calls with attractive costs.

The rest parts of this paper are organized as follows. Section II introduces the architecture overview. Section III presents our implementation of the proposed gateway; Section IV presents the implemented services. Finally, Section V presents our conclusion and perspectives.

## II. ARCHITECTURE OVERVIEW

The objective of our implemented prototype is to provide a gateway based on cellular networks (up to the third generation) and that can be used by users connected over the wired network. An example of services that the gateway provides is the capability to handle messaging functionalities (such as short message service – SMS, multimedia messaging service – MMS) and calls. In order to implement these services in an easy way, the following components are required (Figure 1): the application gateway that can be used by a distant user or application through Web services, the terminal equipment (TE) equivalent to the Data Terminal Equipment-DTE, a terminal adaptor (TA), e.g., a GSM data card (equivalent to the Data Circuit terminating Equipment-DCE) and a mobile equipment (ME) such as a 3G mobile phone (equivalent to the Mobile Station-MS) with its Mobile Termination (MT) that performs radio and network functions (RT and NT respectively).

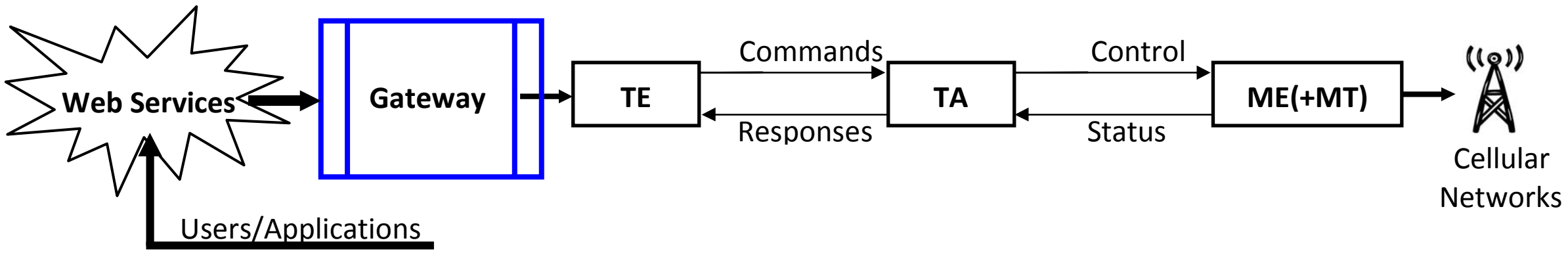

Figure 1. Architecture Overview.

The proposed gateway can be set up on any server machine and remotely used by any authorized user after a required authentication[1]. During the experimentations of our prototype, a cellular modem is connected to the gateway server using a Bluetooth connection. The communication, between the gateway's application and the modem, is also supported using the physical COM port. In order to open the use of the provided services to a large community of users and applications, services are accessible using Web services technology. In addition to the basic services of sending and receiving SMS, MMS and managing calls originated by (or in destination to) the user, the prototype implements, using a set of dedicated modules, other functionalities such as memory data sharing, remote modem's monitoring and a particular application of video surveillance. The interfacing between the different modules that implements the services and the gateway is done using an implemented interface called Mobile Service Agent.

## III. IMPLEMENTATION

In order to experiment our platform, we have implemented a set of services using some GSM/GPRS modems. The modular implementation of the exchange between the cellular modem and the application gateway makes easier to interact with any cellular modem since the modem's control is based on standard AT commands. So any other generations of cellular network such 3G+ or 4G could be easily exploited if we want to enrich the set of supported services. However, during our experimentations, we have noticed that some cellular modems do not recognize some standard commands which implied the use of additional manufacturer commands specific to the modem model. The use of non standard manufacturer commands could be used also for developing advanced services that can not be implemented with standard commands and to optimize some functionalities by new native commands such as supplying efficient and error free TCP/UDP transport mechanisms using the AT interface [7].

### A. Web Services

The prototype implements a SOAP server (using a set of Java Servlets) executed on a Tomcat web server and communicates over the underlying protocol. SOAP is a lightweight protocol, defined by the W3C, intended for exchanging structured information in a decentralized, distributed environment [8]. The latest SOAP W3C recommendation includes, mainly, the definition of an extensible messaging construct that can be exchanged over many underlying protocols (specification's part 1 [9]) and a set of adjuncts that may be used by the messaging framework (specification's part 2 [10]). In our case, the *binding* or carrying SOAP messages with the underlying protocol is based on the HTTP binding in SOAP 1.2-part 2 [10] with respect to the binding specification's discussed in Chapter 4 of [9].

TABLE I. AVERAGE TIME OF ONE SOAP CALL

| Implementation | Average time of one SOAP Call (ms) |
|---|---|
| Web Services (Apache) – AXIS (1.4 final) | 10,10 |
| SOAP Lite (0.65 beta 3) | 28,11 |
| Microsoft SOAP Toolkit 3.0 | 12,22 |
| CORBA (Java Client/Server) | 1,02 |

From the many existing implementations of the HTTP binding, we have tested -on the server side- the implementations presented in Table 1. Tests are based on the Sun Microsystems WSTest [11] and the guidelines of [12] to exclude: the TCP setup time, the time required for getting the response to a Servlet request of Tomcat/Microsoft IIS server, etc. For testing, we have used two hosts on the same switched 100 Mbps Ethernet LAN (only the two hosts are joined via a Cisco Switch): Desktop with a Pentium(R) 4 CPU 2.66 GHz processor and 256 MB RAM using Windows XP Professional, version 2002 - SP3 running as a client. A Laptop with an Intel(R) Core(TM) 2 Duo T7500 2.20 GHz processor and 2 GB RAM using Windows XP Professional 2002 - SP3 running as a server. We adopted the Web Services (Apache)-AXIS implementation [13] within the Tomcat server for its stability, supported features and its easy integration within our prototype.

The interface and modules were implemented using Java APIs in order to perform the serial/Bluetooth communication between the gateway component's server and the cellular modem. Audio stream is handled in the Bluetooth session using the `serial communication class`. Our prototype includes data processing and session opening and managing. Moreover, the prototype is enriched by the capability to read and modify the data exchanged between the cellular modem and the application components and to handle the errors. Theses capabilities are available for all the implemented services supported by the gateway. Note that the use of a Bluetooth communication mode between the gateway application and the modem, requires supporting the driver which is specific to the Bluetooth material in order to initialize the virtual COM port used in the communication exchange from and to the cellular modem.

[1] Security issues are left out of the scope of this paper.

### *B. AT Commands*

Except the initialization and the stream supervision of the communication within the cellular modem, modem operations are performed using a set of AT commands. The two characters "AT" comes from "ATtention"; this abbreviation prefixes all the commands line sent from terminal equipment (TE) to the terminal adaptor (TA). AT commands are used for carrying out several functions related to the mobile equipment capabilities and cellular network services [15][16][17]. The control is achieved through the terminal adaptor. The network services may concerns any kind of services such as GSM/UMTS messaging, fax transmission, voice set-up and operations, wireless communications services, etc. Some of AT commands are defined such way that they can be easily applied to mobile termination of networks other than GSM/UMTS.

AT commands are standardized in the 3rd Generation Partnership Project (3GPP) specifications [16][17] produced by the CT1 working group [18]. The ITU-T Recommendation V.250 [15] includes AT codification and a definition of a format for orderly extension of the AT command set. In order to perform an operation, the prefix AT followed by a command string is sent through the physical/emulated serial port to the mobile's modem and is executed on receipt of carriage return `CR`. A result code is sent from the terminal equipment and indicates the response after the execution of the AT command. For example, the command `AT+CREG?` confirms whether the cellular modem is connected to the cellular network or not. The command `AT+CSQ` indicates the signal strength; it returns received signal strength indication (`RSSI`) and channel bit error rate (`BER`).

As said before, in our system, we have also used additional manufacturer specific commands (SSTK commands). The standard and specific used commands interrogate the cellular modem in order to exploit the cellular network possibilities (services) and achieve a specific action for a service proposed by the gateway. ASCII is the format of command content and each command can return a result code. For example the command `+CHUP` allows answering a call received by the modem and `AT+CGMI` allows retrieving the identifier of the mobile. For the service of managing the calls that a user of the wired network can use through our gateway component, an audio tunnel is used between the TE and our server gateway. This tunnel allows sending and receiving the audio stream from the TE and is transported using the RTP protocol. The gateway components allow the interaction with the cellular modem. For example, our gateway can initiate a call; send textual or multimedia messages using the selected cellular network. To discover and publish the available services available by the mobile terminal node, the gateway can publish available and personalized public services that are implemented. For compatibility reasons, the gateway component extracts all the possible options for a given cellular modem using a standard AT command that gives the list of supported operations of the modem. Available commands can be used in public services that are configured to be shared with other nodes of the network.

To implement a way in which the mobile can communicate with the gateway we were mainly based on the Java Communications API (also known as javax.comm) [19] and Visual Basic API MSCOMM [20]. The first API, used in the major part of our prototype, facilitates developing platform-independent communications applications for technologies such as smart cards, embedded systems, modems, etc. It provides applications access to RS-232 hardware (serial ports) and limited access to IEEE-1284 (parallel ports) in the Standard Parallel Port (SPP). Similarly, MSCOMM handles serial communications for MS environments. The communication Java class handles the communication between the cellular modem and the other modules. It handles the different sessions: opening/closing and managing sessions, read/write methods and listening to a given port. The `communication class` can be used by any independent module in the prototype. An event listener is implemented in order to handle all the extern events that come from the cellular network. Such events could be textual/multimedia messages notifications, calls notifications or other notification regarding personalized services that use the cellular network. The listener implementation is based on the `SerialPortEventListener` interface.

## IV. IMPLEMENTED SERVICES

Our prototype implements the following services:

### *A. Short Message Service (SMS)*

Within a GSM/UMTS mobile terminal, there are three kinds of interface protocols for controlling the SMS functions using an asynchronous interface: Block mode, Text mode and PDU mode [21]. In the Block mode, SMS functions are controlled using binary data in a synchronous way. Each message exchanged between the MT (of the ME) to the TE consists of a data block and block check sum (BCS). Using this mode implies that the control will remain within that mode until the procedure for exiting the mode is executed, after which control is returned to the asynchronous command state. This means that the mobile terminal will not be available for other functions such as voice or data calls until this mode is terminated. Due to theses disadvantages and to the induced complexity, the Block mode is rarely used in practice and is not supported in our implementation.

We have considered both the Text and the PDU mode which are based on AT commands. The Text mode operates in a character-based way: data and SMS parameters are transmitted directly as character strings without the need of additional coding. The PDU mode is based on the same AT commands as the Text mode. It uses the TPDU (Transport Protocol Data Unit) to encode each character in hexadecimal rather than a raw binary format used in the Block mode. In the PDU mode, a complete SMS Message including all header information (such as receiver's number, validity period, etc.) is passed as a binary string while in Text mode the headers are input separately.

Table II presents some AT commands used in the implementation of the short message service. The `+CGMF` command is used to set the input and the output format of the messages to be sent, listed, read or written. The mode parameter indicates the selected mode: Text (`+CGMF=1`) or PDU (`+CGMF=0`). The `+CNMI` command is used to set the used procedure when receiving new messages from the network. The different parameters of this command are used to control and buffer result codes. For instance, when the *mt* variable is set to 0, no `SMS-DELIVER` messages are routed to the TE. The execution command `+CNMA` confirms the correct reception of new messages using a `SMS-DELIVER/SMS-STATUS-REPORT` notification. Sending text messages to cellular numbers is translated from the gateway order into `CMGS` AT commands (`SMS-SUBMIT`). A message reference value us returned to the TE when the message is successfully sent, otherwise (if the delivery fails in the network or the ME), a `+CMS ERROR` is returned.

The text messages translation to AT commands is based on the extraction of the text from SOAP messages. The text content is divided into may short messages with a length of 106 characters. The implemented SMS validity range varies from one hour to a week and the SOAP message can select to save the message in the sending request. The reception of messages is implemented by listening to the COM port of the Mobile Service Agent and the TE link. When a message is received, the TE sends an AT notification to our application then the message is processed.

TABLE II. AT COMMANDS FOR SMS

| AT Command | Function |
| --- | --- |
| +CMGF=[<mode>] | Messages format and mode |
| +CNMI=[<mode>[,<mt>[,<bm>[,<ds>[,<bfr>]]]]] | Procedure of messages' reception |
| +CNMA | Reception confirmation |
| +CMGS | Reception confirmation |

## B. *Multimedia Messaging Service (MMS)*

The Multimedia Messaging Service involves several elements in the overall MMS architecture such as the MMS user agent, MMS Relay/Server, User Databases (that contain user related information such as subscription and profile), Value Added Services (VAS), etc. (see the MMS architecture in Section 4.2 of [22]).

A minimum set of supported formats was defined in the 3GPP standards [22] and [23]: plain text, speech, audio, video, etc. The MMS user agent is responsible of many complex operations such as the user profile management, storing of messages on the terminal, handling of MMS-related information on the (U)SIM, etc. Many of these operations depend to the terminal and the user agent implementation and configuration as it was planned by the standards [22][23]. For instance, regarding the MMS retrieval, messages may be displayed on the terminal with or without any pre-notice. The user could be not aware of the message's notification and storage on the device. These user agent-dependent operations are key functionalities for handling correctly the MMS service of our gateway which make it difficult to bypass the user agent in the case where an MMS capable mobile equipment is used. The same situation occurs if we use a cellular modem (independently to a mobile equipment), in this case we can not bypass the manufacturer toolkit for handling the MMS service.

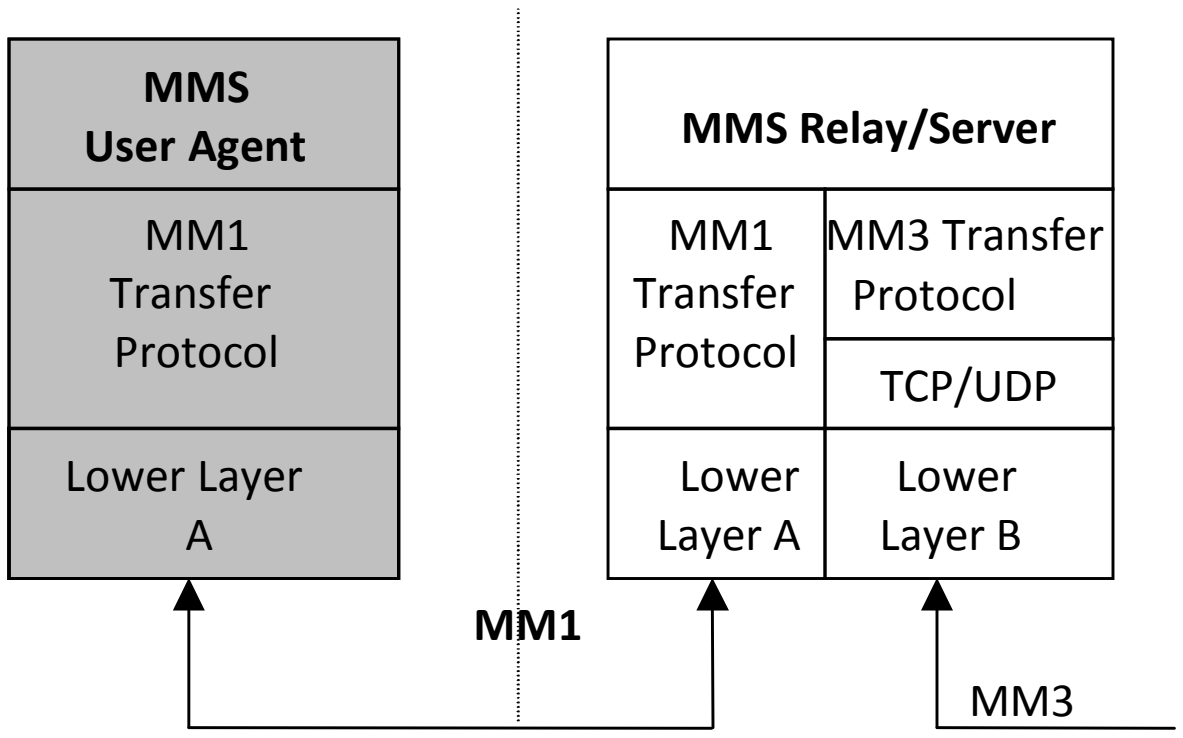

Figure 2. The MM1 Interface.

In this work, the interesting exchange is the one done between the MMS capable user equipment and the MMS environment (MMSE). This exchange is done between the MMS user agent and the MMS Relay/Server. In the MMS reference architecture [22], the concerned interface (reference point) is called MM1 as shown in Figure 2. The MMS functionalities that concern our gateway are the user agent local operations and the MM1 exchange: submission of MMS from the user agent, pulling MMS from the MMS Relay/Server, pushing MMS message from the MMS Relay/Server to the user agent (notification), delivery reports between the MMS Relay/Server and the user agent.

In the implementation of the MMS service non standard AT manufacturer commands are used. For instance, the commands `AT*E2IPA` and `AT*E2IPS` are used within the Sony Ericsson GR47 to activated IP session (based on the stored PDP context) and to receive data with respect to the transmission across TCP/UDP [7]. These commands are combined with others particular commands (such as the `AT+CKPD` and `AT*EKEY`) to force the user agent to perform some needed actions. The considered messages, between the user agent and the MMS Relay/Server, follow a part of the OMA implementation (stage 3) of the MM1 interface [24].

Figure 3 shows these messages in a general scenario. M-Send.req is used to send the MMS, the proxy provides a status code for the requested operation (M-Send.conf). The M-Delivery.ind is used to convey information about the status of a particular delivery -identified by a message-ID - that was performed. When a message is available, the UA is informed by a M-Notification.ind message which let the UA, thanks to the provided information (about location, class, expiration time, etc.), to retrieve the message. The M-NotifyResp.ind provides a message retrieval status code. The UA retrieves the message using a WSP/HTTP GET request. M-retrieve.conf, if successful, contains the message and may request an acknowledgement depending to the MMS Relay/Server needs, e.g., to provide a delivery notice back to the originator of the message or to be able to delete the message from its storage space.

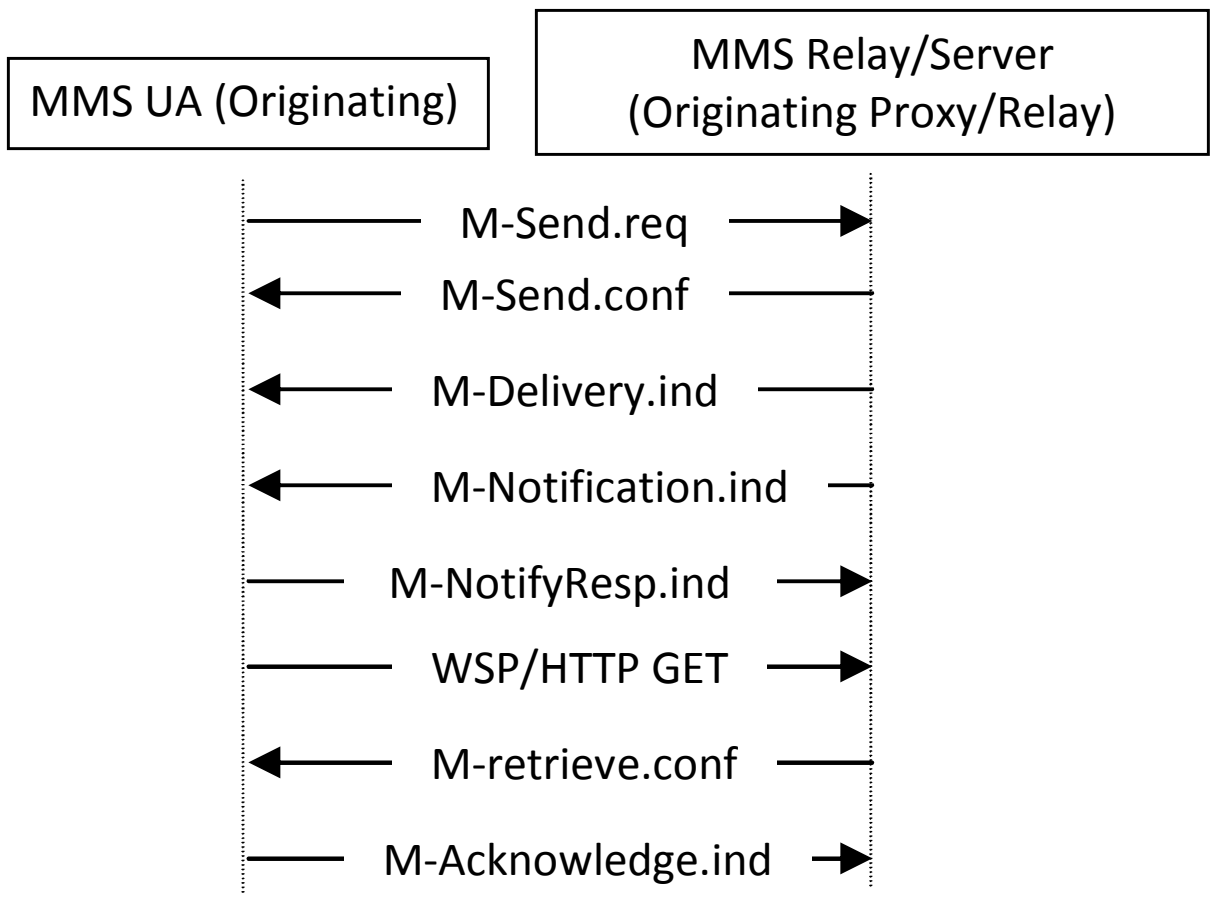

Figure 3. Messages Exchange between the UA and the MMS Relay/Server.

### *C. Call Service and Data Sharing/Synchronization*

Calls and data access are implemented to show the ability to link cellular services to wired users. This can be useful for online service providers or a simple user who aims to use its own mobile terminal to access its data or to take advantages of its mobile subscription offers from anywhere.

As for the SMS service, AT commands are used to perform making and receiving calls based on the cellular modem. The actual state of the work considers only the control of GSM/UMTS voice calls. Data and fax control could be found in the ITU-T recommendations [15]. Some of the related AT commands, used by the gateway, are the +CSTA command to set the type of number for dialing commands and ITU-T V.250 dial command *D* [16] that lists a set of characters for making calls and controlling additional services. The support of additional services relative commands, such as the +CCFC and +CCWA for call forwarding and waiting, is not implemented. The destination number of a call is given as a parameter of the *D* command and can be selected from a memory entry. Also, the destination number can be selected from a phonebook entry using the +CPBR and +CPBS=? commands. To handle incoming calls events (among other events), we use the +CRC=1 command. When an incoming call occurs, an unsolicited result code +CRING: <type> is captured from the COM port. When the type is VOICE [,<priority>[,<subaddr>,<satype>]], this means that the incoming call is a normal voice so it can be processed by the gateway. The +CHUP allows to answer the current received call and the audio stream is exploited by the mean of an audio gateway between the application and the mobile equipment. Sending and receiving the audio stream is done using the audio gateway and streaming the audio between the end user and the gateway uses the Real-Time Transport Protocol (RTP) thanks to the Java class: javax.media.rtp [19].

The user equipment (UE), or mobile station (MS), is considered as the composition of a mobile equipment (ME) and the Universal Subscriber Identity Module ((U)SIM). Data, that the gateway allows to be shared or synchronized, can be stored either at the (U)SIM or the ME level. When a user or a distant application sends an authenticated order to synchronize or share the UE data, this order is translated into a set of AT commands that access to the UE data. The related functionalities includes: accessing and backup its own data (of the UE's owner), sharing these data with other users (other authenticated users could access to the shared folder of the gateway), reading/modifying/finding/removing existing data and hence data synchronization with other external information. Data include the user contacts, mails, SMS, MMS messages, media files (ringtones, pictures, songs, etc.) and call lists. Some of the used commands are: +CPIN required before using the ME, +CSIM and +CRSM for (U)SIM access, +CPBS/CPBR/CPBF/CPBW for managing the phonebook storage.

### *D. Personalized Services: case of a video surveillance*

In addition to basic services such as messaging and calls, the gateway platform is able to provide new personalized services based on the basic operations already implemented. The video surveillance service is an example of a simple service that was implemented within the gateway. It is based on an external movement detection package and a live video stream captured by a camera. When a movement is detected the package sends an order to the wireless gateway in order to send an SMS message to a mobile phone number selected initially. The wireless gateway makes the implementation of this service very easy; the video surveillance application could at any moment change the destination number and way to inform that destination number if a movement is detected. The current implementation of this service sends a simple SMS alert. However, this could be improved, e.g., by sending the captured movement or dangerous presence using the MMS messaging service.

## V. CONCLUSION AND FUTURE WORK

In this paper we have presented a new framework of implementing an easy gateway between users of the wireline network and users of the cellular network. Our gateway has successfully enabled users of the wired network to gain access to a set of typically cellular services such as joining other mobile users using mobile messaging. Another demonstrated service was making mobile terminals data accessible from everywhere and anyhow even using a simple Internet connection. There is scope for lot of improvement depending on the requirements of the gateway's use for instance: the optimization of the QoS regarding the audio streaming, the support of several simultaneous accesses and composing new services based on the implemented modem control and operations. Another point which was not discussed is including some policies regarding the cost sharing, in respect to the cellular operator contract, when a cellular service is used by many users.

To force some particular user agent dependent actions, e.g., to access to its own configuration for the MMS service, we have used emulating commands such as `AT+CKPD` and the Sony Ericsson terminal command `AT*EKEY`. This implementation detail has implied the provision of some mobile terminal profiles to make the use of the MMS service universal and accessible by Web services. The actual profiles repository still small and one of the perspectives is to enrich it with many other models. Also, the SOAP's performance (overhead, execution time, complex message throughput) regarding the transport protocol could be improved. [25] concluded that using UDP as the underlying transport is more suitable than TCP. For us, this can be adopted only for frequent short SOAP messages. In this context, the optimization using Message Transmission Optimization Mechanism (MTOM) and XML-binary Optimized Packaging (XOP) [14] will be explored.